\documentstyle[11pt,newpasp,twoside,epsfig]{article}
\newcommand{\Lya}{Ly-$\alpha$\thinspace} 
 
\newcommand{\eg}{e.g.} 
\newcommand{\ie}{i.e.\thinspace} 
\newcommand{\etal}{et al.\thinspace}

\def\edcomment#1{\iffalse\marginpar{\raggedright\sl#1\/}\else\relax\fi}
\reversemarginpar

\begin{document}
\title{Galaxy Formation Now and Then}
 \author{Matthias Steinmetz}
\affil{Astrophysikalisches Institut Potsdam, An der Sternwarte 16, 14482 Potsdam, Germany}
\affil{Steward Observatory, University of Arizona, 933 N Cherry Ave, Tucson, AZ 85721, USA}

\begin{abstract}

I review the current state of our understanding of the galaxy
formation and evolution process from the modeler's perspective. With
the advent of the cold dark matter model and the support of fast
computers and advanced simulation techniques, there has been
considerable progress in explaining the growth of structure on the
largest scales and in reproducing some of the basic properties of
galaxies and their evolution with redshift. However, many properties
of galaxies are still only poorly understood or appear to be in
conflict with the prediction of the cold dark matter model. I discuss
in what direction the next generation of galaxy formation models may
go and why a large space-based optical-UV telescope could be critical
for the calibration and testing of these advanced models.

\end{abstract}

\section{Introduction}

The past couple of years have witnessed a dramatic increase in the
quantity and quality of observations on the formation and evolution of
galaxies. Galaxies are routinely identified at redshifts exceeding
three and high resolution imaging allows us to study their internal
structure. These data are complemented by high resolution spectroscopy
of QSO absorption systems that provide further clues on the evolution
of baryons in the universe. In fact, this increase has been so rapid
that observations have outgrown their theoretical
framework. Traditional approaches, which rely heavily on the
morphological classification of galaxies and which intend to
disentangle the star formation history of galaxies, seem outdated if
compared with the much richer structure seen in galaxies at different
redshifts.

Motivated by the increasing body of evidence that most of the mass of
the universe consist of invisible ``dark'' matter, and by the particle
physicist's inference that this dark matter consists of exotic
non-baryonic particles, a new and on the long run much more fruitful
approach has been developed: rather than to model the formation and
evolution of galaxies from properties of present day galaxies, it is
attempted to prescribe a set of reasonable initial conditions. The
evolution of galaxies is then modeled based on physical processes that
are considered to be relevant such as gravity, hydrodynamics,
radiative cooling and star formation. The outcome at different epochs
is then confronted against observational data. One scenario that has
been extensively tested in this way is the model of hierarchical
clustering, currently the most successful paradigm of structure
formation.

In this contribution I review the main successes but also some of the
generic problems of models in which structure forms by hierarchical
clustering. I briefly compare the current state of the field with that
10-15 years ago followed by some speculations in which direction the
field may develop in the next decade and how a large optical-UV
telescope in space may support such developments.

\begin{figure*} 
\epsfig{file=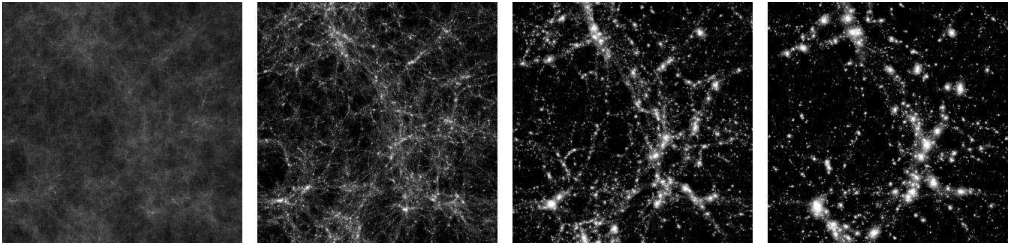,height=3.2cm} 
\caption[]{Time sequence of structure formation in a hierarchical clustering universe,  
here for the so-called $\Lambda$CDM model. The four snapshots correspond (from left to 
right) to redshifts of 9, 3.5, 1 and 0, respectively. The 
simulation box is 50 Mpc (comoving) on the side.} 
\end{figure*} 

\section{The state of the field}

Hierarchical clustering is at present the most successful model for
structure formation in the universe. In this scenario, structure grows
as objects of progressively larger mass merge and collapse to form
newly virialized systems (Figure 1). Probably the best known
representative of this class of models is the {\sl Cold Dark Matter}
(CDM) scenario. The initial conditions consist of the cosmological
parameters ($\Omega, \Omega_{\rm baryon}, \Lambda, H_0$) and of an
initial density fluctuation spectrum such as the CDM spectrum. The remaining
free parameter, the amplitude of these initial fluctuations, is
calibrated by observational data, \eg, the measured anisotropies of
the microwave background. Since the CDM model was introduced in the
early 80s the values of these parameters have been revised and tuned
to match an ever growing list of observational constraints, from the
$\Omega_0=1$, $H_0=50\,$km s$^{-1}$Mpc$^{-1}$, and $\sigma_8=0.6$ of
the former ``standard'' Cold Dark Matter model to the currently
popular ``concordance'' $\Lambda$CDM model.  This $\Lambda$CDM model
envisions an eternally expanding universe with the following
properties (Bahcall
\etal 1999): (i) matter makes up at present less than about a third of the
critical density for closure ($\Omega_0 \approx 0.3$); (ii) a non-zero
cosmological constant restores the flat geometry predicted by most
inflationary models of the early universe
($\Lambda_0=1-\Omega_0\approx 0.7$); (iii) the present rate of
universal expansion is $H_0 \approx 70$ km s$^{-1}$ Mpc$^{-1}$; 
(iv) baryons make up
a very small fraction of the mass of the universe ($\Omega_b \approx
0.04 \ll \Omega_0$); and (v) the present-day {\sl rms} mass
fluctuations on spheres of radius 8 $h^{-1}$ Mpc is of order unity
($\sigma_8 \approx 0.9$).  The hierarchical structure formation
process in this $\Lambda$CDM scenario is illustrated in Figure 1,
which depicts the growth of structure within a $50\,$Mpc box
between redshifts nine and zero.  The $\Lambda$CDM model is consistent
with an impressive array of well-established fundamental observations
such as the age of the universe as measured from the oldest stars, the
extragalactic distance scale as measured by distant Cepheids, the
primordial abundance of the light elements, the baryonic mass fraction
of galaxy clusters, the amplitude of the Cosmic Microwave Background
fluctuations measured by COBE, BOOMERANG, MAXIMA and DASI, the
present-day number density of massive galaxy clusters, the shape and
amplitude of galaxy clustering patterns, the magnitude of large-scale
coherent motions of galaxy systems, and the world geometry inferred
from observations of distant type Ia supernovae, among others.

\begin{figure*} 
\begin{center} 
\psfig{file=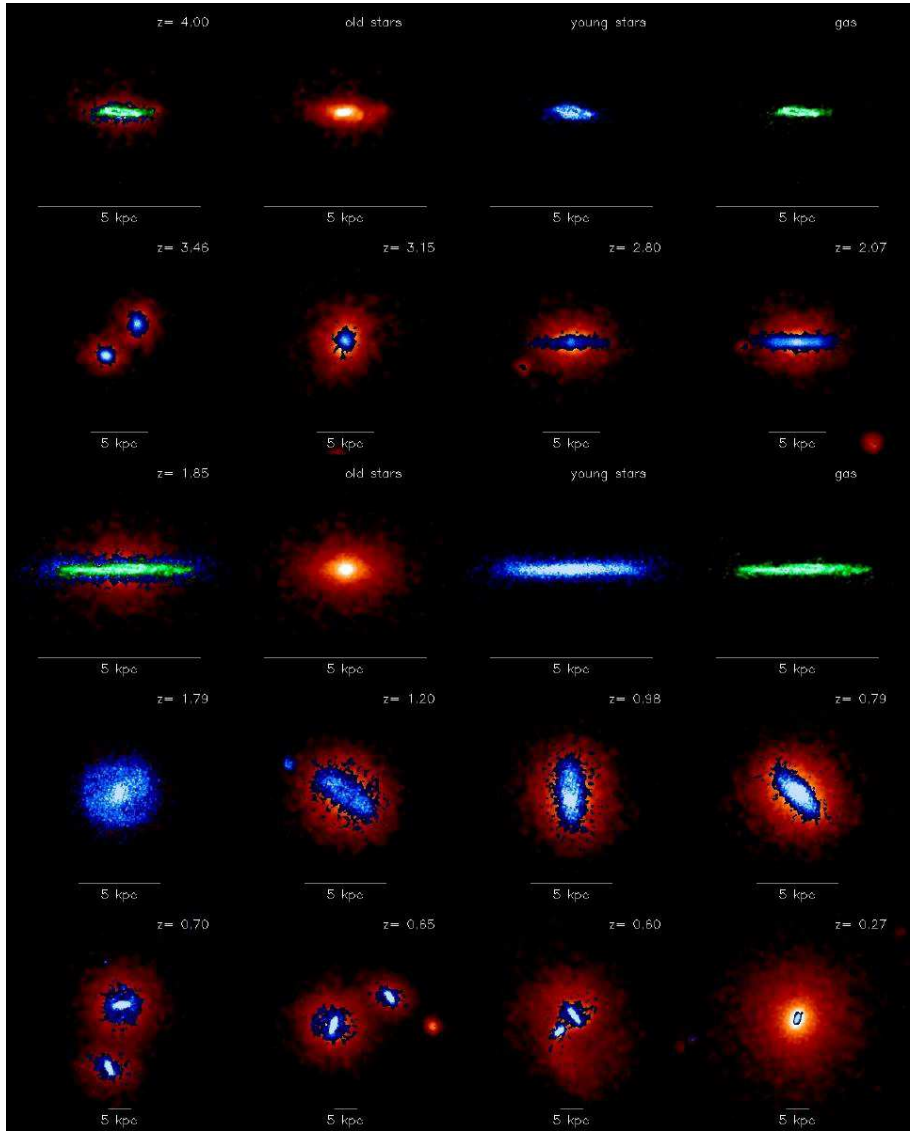,height=15cm} 
\end{center} 
\caption{\label{number} Surface mass density of the gaseous and stellar components of
$\Lambda$CDM   halo at various epochs. Horizontal bars in each panel are 5
(physical) kpc long and indicate the scale of each figure. Rows 2, 4 and 5 show
time sequences near some key evolutionary stage. Rows 1 and 3 decompose a galaxy
at a particular redshift (left) into its constituents: old stars, young stars
and gas (from left to right). Top row: The most massive progenitor at z=4, seen
edge-on. Second row: The formation of a bulge and the rebirth of a disk.  Third
row: The appearance of the galaxy at z=1.8, seen edge-on.  Fourth row: The tidal
triggering of bar instability by a satellite resulting in the emergence of a
rapidly rotating bar. Bottom row: A major merger and the formation of an
elliptical  galaxy.} 
\end{figure*}

The hierarchical build-up is also thought to determine the morphology
of a galaxy, most noticeably the difference between disk--like systems
such as spiral galaxies (some of them barred) and spheroidal systems
such as elliptical galaxies and bulges. This picture envisions that
whenever gas is accreted in a smooth fashion, it settles in
rotationally supported disk-like structures in which gas is slowly
transformed into stars. Mergers, however, convert disks into
spheroids. The Hubble type of a galaxy is thus determined by a
continuing sequence of destruction of disks by mergers, accompanied by
the formation of spheroidal systems, followed by the reassembly of
disks due to smooth accretion (Figure 2). This picture of a
hierarchical origin of galaxy morphology has been schematically
incorporated in so-called semi-analytical galaxy formation models used
to study the evolution of the galaxy population, but its validity in a
cosmological setting has only just recently been directly demonstrated
(Steinmetz \& Navarro 2002).

Numerical simulations have been an integral part in the detailed
analysis of the virtues of the CDM scenario. Only numerical techniques
can account for the highly irregular structure formation process and
for at least some of the complicated interaction between gravity and
other relevant physical processes such as gas dynamical shocks, star
formation and feedback processes. Simulations also provide the
required interface to compare theoretical models with observational data and
are able to link together different epochs. While simulations of
structure formation on the larger scales have mainly used large
massively parallel supercomputers, studies how individual structures
such as galaxies or clusters of galaxies form in the $\Lambda$CDM
scenario have heavily used special purpose hardware like the GRAPE
(=GRAvity PipE) family of hardware N-body integrators (Sugimoto \etal
1990).

Although gas dynamical simulations were considerably successful in
explaining some details of the galaxy formation process, the largest
impact so far has been in the field of QSO absorption
systems. Numerical simulations can reproduce the basic properties of
QSO absorbers covering many orders of magnitude in column density (Cen
\etal 1994; Zhang, Anninos \& Norman 1995; Hernquist
\etal 1996; Haehnelt, Steinmetz \& Rauch 1996). Indeed, gas dynamical 
simulations were even responsible for a paradigm shift, as QSO
absorbers are no longer considered to be caused by individual gas
clouds. Absorbers of different column density (\Lya forest, metal line
systems, Lyman--limit systems and damped
\Lya absorption systems) are rather reflecting different aspects of the
large-scale structure of the universe.  While the lowest column density systems
($\log N \approx 12-14$) arises from gas in voids and sheets of the ``cosmic
web'', systems of higher column density are produced by filaments ($\log N
\approx 14-17$) or even by gas that has cooled and collapsed in virialized halos
($\log N > 17$).

Even though the above list of achievements appears quite impressive, it
mainly addresses structures on scales exceeding a few hundred kpc. On
smaller scales, theoretical models have at best provided some
qualitative insights in the physics of the galaxy formation process,
but we are still far from being able to make quantitative predictions
of the properties of the galaxies now or at higher redshifts on those
small scales. For example, even though we have some qualitative
insight how the Hubble sequence has formed, we are still unable to
account in detail for the mix of morphologies at different redshifts. To
a lesser extent we still do not
have the computing power in order to study the small scale properties
for a cosmologically representative sample of galaxies, but the more
important factor is likely our rather poor understanding of the
astrophysical processes acting on such scales such as star formation,
and energetic feedback from supernovae and stellar winds. 

Furthermore, a list of findings has surfaced in the past few years that seem to be at
odds with the model predictions: (i) Because of their negligible
primordial velocity dispersion, cold dark matter particles can achieve
enormous phase space densities. As a result, numerical simulations
have consistently shown that near the centers of halos the density
profiles of virialized CDM halos diverge as $r^{-1}$ (Navarro, Frenk
\& White 1996) or perhaps even as steeply as $r^{-3/2}$ (Moore \etal
1999a). These divergent profiles are at odds with the usual
interpretation given to the ``solid-body'' HI rotation curves reported
for some low surface brightness (LSB) dwarf galaxies (Flores and
Primack 1994, Moore 1994). (ii) Another generic prediction of CDM  models
is that virialized galactic halos will typically be triaxial and have
dense cores. Yet in a number of galaxies where detailed stellar and
gas dynamical observations are possible, the dark matter contribution
within the optical radius appears to be rather small: many disks, perhaps including
the Milky Way, are ``maximal'' (see, e.g., Debattista
\& Selwood 1999). (iii) High resolution N-body simulations indicate
that, if the dark halo of our own Milky Way is made up of CDM, it
should contain several hundred dark matter sub-condensations; a number
that climbs to roughly one thousand if all halos within the Local
Group are considered (Klypin \etal 1999, Moore \etal 1999b).  On the
other hand, observations of the Milky Way surroundings and of the
Local Group reveal an order of magnitude fewer galaxies than expected in this picture
(Mateo 1998). (iv) A difficulty indirectly related to the substructure
problem concerns the angular momentum of gaseous disks assembled in
hierarchical clustering scenarios. In the absence of heating, most
of the mass of a galactic disk forming within a CDM halo is accreted
through mergers of proto galaxies whose own gas components have
previously collapsed to form centrifugally supported disks. Numerical
simulations (Navarro \& Benz 1991, Navarro, Frenk \& White 1995,
Navarro \& Steinmetz 1997) show that most of the angular momentum of
the gas is transferred to the surrounding halos during mergers. As a
result, the spin of gaseous disks formed by hierarchical mergers is
much lower than those of observed spirals.

It should be clear from this list that a considerable concerted effort
is required by the community in order to promote our cosmological
concordance model to a concordance model of galaxy
formation. Considering the complexity of the processes involved,
progress is likely to continue to be drive by observations.

\section{Where did we come from ?}

Before speculating what the state of the field may be like in 10--15
years, it may be illustrative to reconsider how the current state
described above compares to the state of the field 10--15 years ago,
i.e. before the advent of the Hubble Space Telescope and ground-based
8m class telescopes. Cosmological parameters were only very poorly
constrained, the Hubble constant was known to only within a factor of
2, $\Omega$ at best to within a factor of ten. Hierarchical clustering
in the form of a $\Omega=1$ CDM model was already favored by most theorists, 
but it was certainly not accepted within the larger community as it is
today. Similar statements can be made concerning the more basic concept that
the morphological type of a galaxy may primarily reflect its merging history. From the
observational side, barring some episodic evidence of very high
redshift objects (radio galaxies, QSOs) and some indirect evidence for
evolution (e.g.~Butcher-Oemler effect), whose relationship to regular
galaxies was quite unclear, galaxy properties were mainly determined
only at very low redshift. In fact, most galaxy models did not, or if so
only very schematically, include a cosmological context. Similarly,
galaxy populations were mainly discussed in terms of non-evolution {\sl vs}
passive evolution models. It should be clear from this listing how
rapidly the field of galaxy formation has changed in the past decade.

\section{Prospects}

Considering this rapid progress it seems (and probably is) impossible
to make sound predictions how the field may change over the next ten
or even fifteen years. Nevertheless, the exercise may be entertaining
especially if one comes back for an evaluation after ten
years. Assuming that the field continues to progress at the same pace
and that basic concepts (like $\Lambda$CDM as a cosmological
concordance model) remain unchanged, the following list appears to be
``fair bets'':

\begin{itemize}
\item Microwave background experiments, in particular satellite missions 
like MAP and PLANCK, will accurately determine the cosmological
parameters. Extragalactic astronomy will therefore be dominated by the
quest to understand the formation and evolution of galaxies and the
astrophysics behind it, and will no longer be just a tool to measure
cosmological parameters.
\item The Next Generation Space Telescope (NGST) will provide us 
with a look at the so-called dark ages of the universe, \ie the
universe at redshift larger than six. Furthermore, NGST will allow us to study the
evolution of the stellar component between redshift 3 and 1, the epoch
in which galaxies appear to have developed from some irregular bright clumps
into the Hubble sequence.
\item Astrometric missions like GAIA will provide a detailed dynamical and 
chemical record of how the Milky Way has been built up and how it
transformed its gas into stars.
\item In ground-based astronomy, the second generation of instrumentation on 
8m-class telescopes will be routinely used, and (from an optimist's point of view) 
first data may already come in from 30m optical telescopes.
\item Galaxy formation models will largely benefit from the increase in 
computing power. According to Moore's law, the speed of computers
doubles every 18 month. If this trend continues to hold, then standard
PCs in 15 years will have a speed comparable to the largest
supercomputers today. By then numerical simulation should be capable
of simulating cosmologically representative volumes (a few hundred
Mpc) of the universe at a resolution better than a kpc. The understanding
of astrophysics rather than the availability of sufficient CPU power will
limit our insight into the details of the galaxy formation process.

\end{itemize}

What will be the role of a 8m optical/UV space telescope in this
framework ? Two potential applications that depend critically on the
availability of large UV-sensitive instruments are:

\begin{itemize} 
\item {\sl Observe the development of the Hubble sequence.} Deep high-resolution 
imaging of UV dropout galaxies between redshift 3 and 1 will unravel
the building blocks of typical L$^*$ galaxies at the present day
epoch. Furthermore the transition from clumps at $z\approx 3$ to
Hubble-type galaxies at $z\approx 1$ can be observed. By low-resolution spectroscopy,
the kinematics and thus the development of scaling
relations like the fundamental plane or the Tully-Fisher relation can
be investigated.
\item {\sl Create a 3D-map of the baryons in the universe between z=3 and z=0.} 
Using fainter QSOs (as obtained, \eg, by the Sloan Survey) as background
sources, several patches of the night sky can be probed by many
lines-of-sight. High-resolution spectroscopy of these background
sources will map out the distribution of baryons in the IGM and
eventually provide detailed 3D maps of the gas density, temperature,
metallicity and ionization state. The current paradigm that galaxies
preferentially form at the intersections of filaments in the cosmic
web can be directly probed. Studies of the transverse proximity effect
measure in detail the evolution of the cosmic UV background and thus
on the reionziation history of the universe. Furthermore they can
constrain the amount of beaming in the UV flux of QSOs.
\end{itemize} 

Should these developments be accompanied by a corresponding increase in our
understanding of how stars form, how they release energy to the ISM
and thus how star formation interacts with the evolution of galaxies, 
we may indeed make a major step toward a standard model of galaxy
formation by 2020.

\end{document}